\documentclass[aps,prc,superscriptaddress, preprintnumbers, nofootinbib,10pt, showpacs, floatfix]{revtex4-1}

\usepackage{multirow}

\usepackage{feynmf}

\usepackage{amsmath,amssymb}

\usepackage{graphicx,bm}

\usepackage{slashed}

\usepackage{epstopdf}

\usepackage{ulem} %% for strike-through

\usepackage[usenames]{color}

\usepackage{float}

\renewcommand\sout{\bgroup \color{red} \ULdepth=-.5ex \ULset}

\begin{document}
\title{Dynamical Derivation of the Momentum Space Shell Structure for Quarkyonic Matter}

\author{Kie Sang Jeong}
\affiliation{Asia Pacific Center for Theoretical Physics, Pohang, Gyeongbuk 37673, Republic of Korea}
\affiliation{Institute for Nuclear Theory, University of Washington, Seattle, WA, 98195, USA}
\author{Larry McLerran}
\affiliation{Institute for Nuclear Theory, University of Washington, Seattle, WA, 98195, USA}
%\author{Sanjay Reddy}
%\affiliation{Institute for Nuclear Theory, University of Washington, Seattle, WA, 98195, USA}
\author{Srimoyee Sen}
\affiliation{Institute for Nuclear Theory, University of Washington, Seattle, WA, 98195, USA}
%\email{ksjeong@uw.edu}

\date{\today}

\preprint{INT-PUB-19-048, APCTP Pre 2019-020}

\begin{abstract}
The phase space structure of zero temperature Quarkyonic Matter is a Fermi sphere  of Quark Matter,
surrounded by a shell of Nucleonic Matter.  We construct a quasi particle model of Quarkyonic Matter based on the constituent quark model, where the quark and nucleon masses are related by $m_Q = m_N/N_c$, and $N_c$ is the number of quark colors.  The region
of occupied states is for quarks $k_Q < k_F/N_c$, and for nucleons $k_F < k_N < k_F + \Delta$.  We first consider the general
problem of Quarkyonic Matter with hard core nucleon interactions.  We then specialize to a quasi-particle model where the hard
core nucleon interactions are accounted for by an excluded volume.  In this model, we show that the nucleonic shell forms past some
critical density related to the hard core size, and for large densities becomes a thin shell.  We explore the basic features of such a model,
and argue this model has the semi-quantitative behaviour needed to describe neutron stars.
\end{abstract} 
\maketitle

\section{Introduction}

The concept of Quarkyonic Matter was introduced to address the following paradox that arises in the large number of colors, $N_c$, limit of QCD~\cite{McLerran:2007qj}:  When the nucleon density, $n_ N$ of zero temperature matter is parametrically large compared to the QCD scale, $n_N  >> \Lambda_{\textrm{QCD}}^3$, we would naively expect that the interactions of quarks would be weak, and we would transition to an almost free gas
of quarks.  On the other hand the confining potential of QCD is at finite density attenuated only by particle-hole pairs of quarks, and these
contributions are suppressed by $1/N_c$.  In the large $N_c$ limit, the Debye screening length therefore diverges, $r_{\textrm{Debye}} \sim \sqrt{N_c}  \rightarrow \infty$ ~\cite{McLerran:2007qj} .   Matter is confined and we expect that the effect of interactions is large.

The apparent paradox can be resolved in the following way:  There is a Fermi sea of quarks formed at asymptotic density that may be treated as an almost free gas of quarks.  Near the Fermi surface, long distance interactions of quarks are important. Quark  excitations can bind into nucleons and condense to make a Fermi surface of nucleons.  Particle hole excitations near this Fermi surface are bound into mesons. The  system's essential feature is it  is  a Fermi sea of quarks surrounded by a Fermi shell of nucleons.  Such a description is expected to be valid until parametrically high baryon densities where the Debye screening length is of order the QCD scale, $k_F \sim \sqrt{N_c} \Lambda_{\textrm{QCD}}$ ~\cite{McLerran:2007qj}.  If one includes the effect of finite temperature, the thermal excitations would be mesons and glueballs.

The concept of Quarkyonic Matter leads to a fundamentally new way to conceptualize a transition from Nucleonic Matter to Quark Matter.
In the traditional way of thinking one imagines that there are two different phases of matter, and a phase transition between them.  Either one has entirely Quark Matter or Nucleonic Matter.  In  first order phase transitions, there can of course be different regions of space in these different phases, but in each of the regions it is either entirely Quark Matter or Nucleonic Matter.  For a first order phase transition,
this will lead to a soft equation of state, since in the region of the mixed phase, the pressure remains constant while the energy density changes \cite{DiToro:2009ig, Hebeler:2013nza}.

The transition for Quarkyonic Matter involves a mixed phase in momentum space.  There is no requirement that the pressure
be the same from the two separated regions of nucleons and quarks, and therefore as the energy density changes corresponding to different portions of quarks and nucleons, the pressure will change.  Therefore, Quarkyonic Matter need not have a soft equation of state.   

One might object that there need be a phase transition between Nuclear Matter and Quarkyonic Matter, but even this need not be the case, since quarks and nucleon degrees of freedom are dual to one another. In fact such a description was illustrated in \cite{Fukushima:2015bda}. In effect, the treatment of Quarkyonic Matter is simply realizing different
approximations for the same degrees of freedom corresponding to different kinematic regions.  Of course a true phase transition
would occur at finite temperature, since at large $N_c$, the transition between lower but finite $T$ matter, which has no baryons in it because
the baryon mass is of order $(N_c m_Q) \sim N_c \Lambda_{\textrm{QCD}}$, and baryonic matter that has baryons by virtue of a larger Fermi energy, $E_F \ge N_c \Lambda_{\textrm{QCD}}$ ~\cite{McLerran:2007qj}.  Here we are however considering matter that is at very low $T$ and has a finite baryon number density.  We cannot of course rule out the possibility that there may be a very weak first order or second order phase transitions at zero $T$ and finite density, but  the existence of Quarkyonic matter by itself does not require this in the region where the baryon density is $n^N \ge \Lambda_{\textrm{QCD}}^3$.  It is amusing to note that the Fermi momenta of ordinary nuclear matter for $N_c = 3$ is of order $k_F \sim \Lambda_{\textrm{QCD}}$, so that the density
of nuclear matter is parametrically of order $\Lambda_{\textrm{QCD}}^3$

In the large $N_c$ limit, it is very easy to see that the transition from Nucleonic Matter to Quarkyonic matter can generate a very hard equation of state.  This was seen explicitly in a model constructed in Ref.~\cite{McLerran:2018hbz} and was followed up in \cite{Philipsen:2019qqm, Han:2019bub}.  For ordinary nuclear matter, $k_F \sim \Lambda_{\textrm{QCD}}$.  In the additive quark parton model this would correspond to a very small Fermi momenta for quarks, $k_Q \sim k_F/N_c$.  The density of nucleons is initially of order $\Lambda_{\textrm{QCD}}^3$.  On the other hand the baryon  density of quarks is very small $n_Q \sim \Lambda_{\textrm{QCD}}^3 /N_c^3$.  If we change the density by of order one by introducing quarks, then
the change in the quark density implies that $k_Q \rightarrow \Lambda_{\textrm{QCD}}$.  This means changing the baryon density associated with quarks by of order one generates an order $N_c$ change in Fermi momenta for the quark and a corresponding rapid increase in the pressure.
The degrees of freedom are rapidly changing from non-relativistic to relativistic degrees of  freedom.  

This hardness of the equation of state is precisely what is needed to describe the properties of neutron stars \cite{Demorest:2010bx, Antoniadis:2013pzd}.  In a number of analyses, it was argued that at a few times nuclear density the sound velocity becomes of the order of or exceeds $v_s^2 \ge 1/3$ (measuring sound velocity in units of the velocity of light.)~\cite{Masuda:2012ed, Kojo:2014rca, Bedaque:2014sqa, Ma:2018qkg, Tews:2018kmu, Vuorinen:2018qzx, Fujimoto:2019hxv},  The interpolation procedure advocated by Kojo and colleagues~\cite{Kojo:2014rca}, is a procedure that can be motivated from the Quarkyonic Matter hypothesis.

In this paper, we consider the effect of hard core nucleonic interactions on the formation of a shell of Nucleonic Matter within 
Quarkyonic Matter.  In the next section we consider a set of simple mean field models composed of a non-interacting Fermi sphere of quarks
and gas of nucleons with hard core interactions.  We argue that a maximum in the sound velocity signals the transition from Nucleonic Matter to Quarkyonic Matter.  In the following section we discuss how to implement such hard core interactions for
nucleons as an excluded volume gas of free nucleons, where the volume excluded is the hard core  of nucleons.  We next construct a quasi particle model of an excluded volume gas of nucleons on a Fermi shell together with a gas of quarks within the additive quark parton model.  In this model, we show that at low density, matter is composed of nucleons, but makes a transition to Quarkyonic Matter at some density of order the hard core density.  We  compute the thickness of the shell, and show that it is a thin shell when the total baryon density of quarks and nucleons is large compared to the hard core density.  Finally, we compute the sound velocity as a function of baryon number density.

\section{Mean Field Model}
\subsection{The Nucleon Contribution}

We first consider nucleonic matter.
We postulate a relationship between the baryon number chemical potential and the density:
\begin{equation}
   \mu_N - M = \kappa {M \over N_c^2}  \left\{ (1-n_N^N/n_0)^{-\gamma} -1 \right\}.
\end{equation}
In this equation, the baryon number chemical potential is $\mu_N$, the nucleon mass is $M$, $N_c$ is the number of colors, $n_N^N$ is the baryon density contained in nucleons, $n_0$ is the hard core limiting density of this theory, and there are two dimensionless parameters $\gamma$ and $\kappa$
which need to be determined phenomenologically.  The $-1$ term on the right hand side of this equation guarantees that $\mu_N \rightarrow M$ 
as $n_N^N \rightarrow 0$.  (If we were to use this expression at densities of the order of that of nuclear matter, we would no doubt want to include low density corrections in order to match onto previous nuclear matter calculations.)

The quantity $\mu_N -M$ measures the kinetic energy plus interaction energies of nucleons.  The parameter $n_0$ is the density at which this energy diverges, corresponding to a limiting hard core density of nucleons.  In reality, in any realistic model, this should get cutoff when this energy is of the order of the nucleon mass, or when $n_0 - n_N^N \sim 1/N_c^{2/\gamma}$.  We will take $n_0$ to be generically of the order of a few times the density of nuclear matter.  This limiting value corresponds to the ordinary large $N_c$ counting for nucleon interactions.  However at densities corresponding to nuclear matter densities, the Fermi momenta of nucleons is of order $\Lambda_{\textrm{QCD}}$, so that kinetic energies and interaction energies are of order $\Lambda^2/M \sim \Lambda/N_c \sim M/N_c^2$, so this expression has the correct behavior when
$n_0 - n_N^N \sim n_0$.

We are going to be interested in this model for densities approaching the hard core density $n_0$ which is where the expression begins to get singular.  The singularity at this density will eventually be tempered by the inclusion of quark degrees of freedom.  We will do this in the spirit of Quarkyonic Matter where a quark contribution is included additively to the nucleon contribution.  We will eventually discuss how to do this in a way consistent with the Pauli exclusion principle, but this comes later.

At low densities we could modify our model with a power of $n_N^N$ so that we could get the proper low density limit corresponding to a non-relativistic ideal Fermi gas, but for the generic considerations here, we will not make this modification.  Note that at low densities in this theory
\begin{equation}
 \textrm{lim}_{n \rightarrow 0} ~(\mu_N-M) \sim \kappa \gamma {M \over N_c^2} {n_N^N \over n_0}.
\end{equation}
This gives a sound velocity at low density of
\begin{equation}
  v_s^2 = {n_N^N \over {\mu_N~ d n_N^N / d\mu_N}} \sim {\kappa \gamma \over N_c^2} {n_N^N \over n_0}.
\end{equation}
If we take a $\kappa \gamma \sim 1$ for $n_N^N /n_0 \sim 0.25$ this gives a reasonable sound velocity of $v_s^2 \sim 0.03$.
Some tuning of $\kappa$ can of course be done to match this on to a more reasonable low density equation of state.

This relationship between $\mu$ and $n_N^N$ has a singularity at $n_N^N \rightarrow n_0$.  It is reminiscent of the relationship the entropy and temperature in Hagedorn models,
\begin{equation}
       S/S_0 \sim (T_c - T)^{-\alpha}.
\end{equation}
where $S$ is the entropy and $T$ the temperature. There is one big difference:  the relationship of the extensive quantities $n_N^N$ to the intensive quantity $\mu_N$ is reversed relative to that of the extensive $S$ intensive $T$.  This reversal is essential. The Hagedorn model, 
can be justified at large $N_c$. When quark and gluon degrees of freedom are included the Hagedorn limiting temperature plays the role of a phase transition temperature corresponding to the entropy density becoming of order $N_c^2 T^3$.  In our case there is no phase transition, but in fact a rapid change in $\mu_N -M$ going from order $1/N_c$ to of order $N_c$.  This is precisely the opposite limit for a phase transition
since the pressure and chemical potential rise rapidly while the density approaches a constant hard core density  (The energy density for our case does not change so much since it is already dominated by the nucleon mass.).  When quark degrees of freedom are included we will see that this behavior arises as one makes a transition to a quark Fermi gas.  

We can construct the energy density using the thermodynamic relationship
\begin{equation}
  {{d\epsilon_N^N} \over {dn_N^N}} = \mu_N  = M + \kappa {M \over N_c^2}  \left\{ (1-n_N^N/n_0)^{-\gamma} -1 \right\}.
\end{equation}  
or
\begin{equation}
   {{\epsilon_N^N - Mn_N^N} \over {n_0 M}} = {\kappa \over {(\gamma-1) N_c^2}}  \left\{ (1-n_N^N/n_0)^{1-\gamma} +(1-\gamma)n_N^N/n_0  \right\}.
\end{equation}

We can construct the pressure from
\begin{equation}
{{dP_N^N} \over {d\mu_N}} = n_N^N.
\end{equation}
Using
\begin{equation}
   {n_N^N \over n_0} = 1- \left\{      {{(\mu_N-M) N_c^2} \over {\kappa M}}     +1           \right\}^{-1/\gamma},
\end{equation}
the sound velocity is therefore (ignoring a small term that vanishes in large $N_c$)
\begin{equation}
v_s^2 = \gamma {{n_N^N/n_0} \over {1 - n_N^N/n_0}} \left\{1 + {N_c^2 \over \kappa} (1-n_N^N/n_0)^\gamma \right\}^{-1}.
\end{equation}
Note that the sound velocity diverges as one approaches the hard core density.  This will be cured by adding quarks.

Another interesting feature of this mean field theory is that the chemical potential, $\mu_N$ is more singular than 
the energy per particle as one approaches the hard core density.  This is because the nucleon density is saturating, and the
thermodynamic identity $d\epsilon/dn = \mu$.  It seems that this observation would make it difficult to realize a quasi-particle
model with this singularity.  In the next section we will show how this is resolved in an excluded volume theory of nucleon interactions.
In an excluded volume theory, one imagines nucleons have hard cores, and that they propagate freely in the volume exclusive 
of that occupied by the hard cores of nucleons~\cite{Zalewski:2015yea, Redlich:2016dpb, Vovchenko:2015vxa}.

\subsection{Including a Contribution from Quarks}

Let us add a contribution to the energy density from quarks so that
\begin{equation}
\epsilon^N = \epsilon_N^N(n_N^N) + \epsilon_Q^N(n_Q^N),
\end{equation}
where the nucleon contribution to the nucleon density $n^N_N$ plus that of quarks $n_Q^N$, satisfies
\begin{equation}
   n^N = n_N^N + n_Q^N.
\end{equation}
Minimizing the energy with respect to $n_Q^N$ or $n_N^N$ at fixed $n_N$ gives
\begin{equation}
  {{d\epsilon_Q^N} \over {dn^n_Q}} = {{d\epsilon_N^N} \over {dn^N_N}}.
\end{equation}
Using the general thermodynamic relation between chemical potnetial and derivative of energy density with respect to density
\begin{equation}
  {{d\epsilon^N} \over {dn^N}} = \mu. 
\end{equation}
Let us define
\begin{equation}
  {{d\epsilon_Q^N} \over {dn_Q^N}} = {{d\epsilon_N^N} \over {dn_N^N}} \equiv \mu_N.
\end{equation}
We can now use the minimization condition to relate $\mu_N$ and $\mu$ as follows
\begin{eqnarray}
\mu=\frac{d\epsilon^N}{dn^N}&=&\frac{d\epsilon^N_N}{dn^N_N}\frac{dn^N_N}{dn^N}+\frac{\epsilon_Q^N}{dn^N_Q}\frac{dn^N_Q}{dn^N}\nonumber\\
&=&\frac{d\epsilon^N_N}{dn^N_N}\left(1-\frac{dn^N_Q}{dn^N}\right)+\frac{d\epsilon_Q^N}{dn_Q^N}\frac{dn^N_Q}{dn^N}\nonumber\\
&=&\frac{d\epsilon^N_N}{dn^N_N}=\mu_N.
\end{eqnarray}
Therefore the energy density, density and pressure from both the quarks and the nucleon should be evaluated at equal chemical potential and is simply the sum of two terms: one from quarks and one from nucleons.
The sound velocity is 
\begin{equation}
  v_s^2 = {{n_N^N + n_Q^N} \over {\mu_N (dn_N^N/d\mu_N + dn_Q^N/d\mu_N)}}.
\end{equation}
Note that at nuclear matter baryon densities, the quark contribution to the energy density and baryon number density will be of order
$1/N_c^{3}$ and in the spirit of the large $N_c$ expansion may be ignored. The physics is totally dominated by nucleon degrees of freedom,
and the energy density and sound velocity are determined by that of ordinary nuclear matter.
On the other hand, at high densities when $n_N^N \rightarrow n_0$, then the degrees of freedom are those of quarks, and will be the result of quark matter computations.

We have computed the sound velocity for mean field equations of state of the type above for typical values of $\gamma \sim 1$ and $\kappa \sim 1$,  The generic result is that at densities much larger $n^N >> n_0$, the sound velocity approaches $1/3$.  It rapidly rises from a very small value to a value of order $1$ at densities somewhat below that of the hard core density.  For large values of $\gamma$ and $\kappa$,
the sound velocity approaches $1/3$ from below, and for values of order 1  it exceeds $1/3$ and approaches it from above with a maximum
near the hard core density.  For very small values of $\gamma$ and $\kappa$, the sound velocity at maximum exceeds 1, and these models
are un-physical.  This is presumably an artifact of treating nucleons using a singular hard core interaction.
 
\begin{figure}
\includegraphics[width=0.45\textwidth]{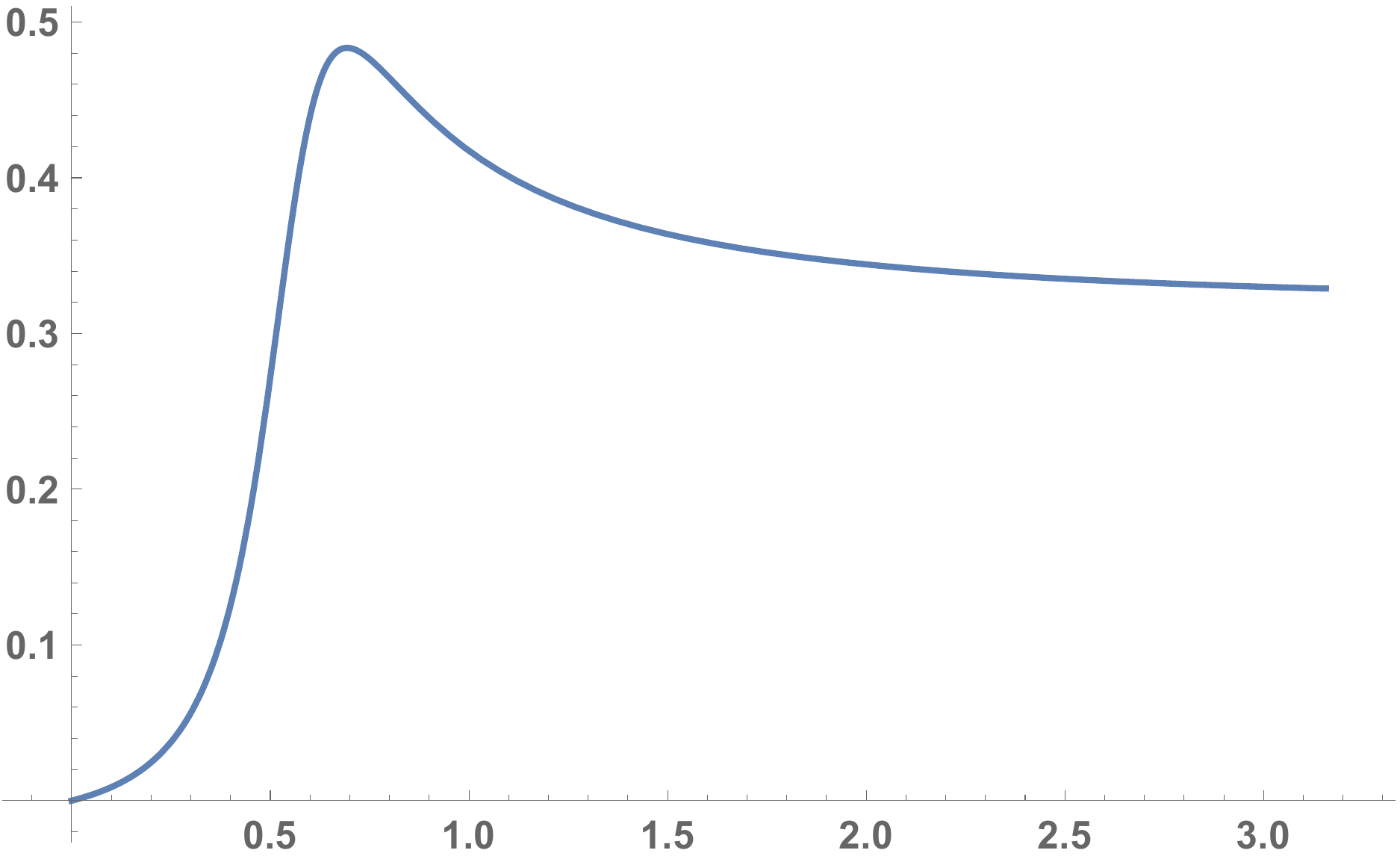}
\caption{A plot of the sound velocity vs the baryonic density of matter.  We have chosen $\gamma = 0.7$ and  hard core density $n_0^N$
to be $0.6~\textrm{fm}^{-3}$} \label{meanfield}
\end{figure}

 In Fig.~\ref{meanfield}, we plot the sound velocity as a function of the total baryon density for the case $\gamma = 1$ and with hard core density $n_0 = 0.6~\textrm{fm}^{-3}$.  This is only an example, and if one wanted to use the mean field equations of state to describe properties 
 of neutron stars, tuning would certainly be needed.  Note the characteristic maximum in the sound velocity that occurs near the hard core density
 as matter transitions from dominantly nuclear matter to a mixture quark matter and nuclear matter.   

Now in reality, the situation will be more complicated.  In these considerations, we have ignored interactions between nucleons and quarks,
and such simple results are not obtained.  The equation of state for quarks also for low quark density should show significant deviation from an ideal gas of constituent quarks.  On the other hand, we will see that ignoring such possible interactions leads to a simple model with the characteristics needed for the phenomenology of the equations of state that describe 
neutron stars.

\section{An Excluded Volume Theory of Nuclear Interactions:  Nucleons Only}

In this section, we show how to construct an excluded volume theory of nuclear interactions~\cite{Zalewski:2015yea, Redlich:2016dpb, Vovchenko:2015vxa}.  In this section,we will consider a theory of nucleons only, and our purpose is to show how such a theory generates a singular chemical potential near the hard core density, and also how
the theory is consistent with various thermodynamic identities. In a later section we include quarks.

If we have  nucleons with a hard core radius $r_0$, and a hard core volume $v_0 = {4 \over 3} \pi r_0^3$, then we can define the  hard core density as
\begin{equation}
  n_0 = 1/v_0.
\end{equation}
For a system with baryon density $n$ and volume $V$, the excluded volume not occupied by baryon cores is
\begin{equation}
   V_{ex} = V~ (1 - n/n_0).
\end{equation}

We assume the nucleons are free particles within the excluded volume so that
\begin{equation}
  n_{ex}= {n \over {1-n/n_0}} = {2 \over {(2\pi)^3}} \int^{k_F} d^3p.
\end{equation}
Here we have assumed two spins and one flavor of nucleons.  The energy density of matter in the excluded volume is
\begin{equation}
  \epsilon_{ex} = {2 \over {(2\pi)^3}} \int^{k_F} E_p ~d^3p,
\end{equation}
where $E_p = \sqrt{p^2 + M^2}$.  The energy density and the excluded volume energy density are related by
\begin{equation}
  \epsilon = (1 - n/n_0) \epsilon_{ex}.
\end{equation}
The pressure is
\begin{equation}
  P = - {{dE} \over {dV}},
\end{equation}
where $E$ is the total energy and $V$ is the total volume,
\begin{equation}
   P = -\epsilon  + n {{d\epsilon} \over {dn}}.
\end{equation}

Now define a chemical potential
\begin{equation}
  \mu  = {{d\epsilon} \over {dn}},
\end{equation}
and an excluded volume chemical potential
\begin{equation}
 \mu_{ex}  = {{d\epsilon_{ex}} \over {dn_{ex}}}.
\end{equation}

One can use these identities to verify that the pressure also can be expressed in terms of excluded volume quantities as

\begin{equation}
  P = -\epsilon_{ex} + \mu_{ex} n_{ex}.
\end{equation}

The equality of the two expressions for the pressure require that
\begin{equation}
    \mu = {{\mu_{ex}  - \epsilon/n_0} \over {1-n/n_0}}\label{chm1}.
\end{equation}
This can also be written as 
\begin{equation}
  \mu -M = {{\mu_{ex}-M - (\epsilon-Mn)/n_0} \over {1-n/n_0}}.
\end{equation}

We can now understand the singularities we had in mean field theory.  For example, if the system is non-relativistic, 
$\epsilon - Mn \sim k_{ex}^{~2} \sim (1-n/n_0)^{-2/3}$, $\mu_{ex} -M \sim k_F^2$ while the true chemical potential is singular as
$\mu \sim (1-n/n_0)^{-5/3}$.  This is precisely what we observed in mean field theory that the chemical potential is one power
more singular as $n \rightarrow n_0$ than the energy density.  It is reassuring to see this occur in our simple quasiparticle model.
In the next section, we generalize our excluded volume computation to put quarks in a Fermi sea and nucleon in a shell
of Fermi momenta outside the Fermi sea.  This modifies the exponent of the dependence upon density.

\section{Shell Equations}

In this section, we consider nuclear matter in a shell with momenta greater than that of $N_c k_Q$ wher $k_Q$ is the quark Fermi momentum.
Recall that in the additive quark model, the quark Fermi momenta and nucleon Fermi momenta are  related by $k_F = N_c k_Q$.  Our physical picture will be that below the shell of nucleons there is a filled Fermi sea of quarks.

We let the excluded volume density be
\begin{equation}
  n^N_{ex} = {{n_N^N} \over {1 -n_N^N/n_0}}.
\end{equation}
Here $n^N_{ex}$ is the excluded volume nucleon density, and $n_N^N$ the nucleon density.  The hard core
density is $n_0$
From this we compute for a shell of nucleons:
\begin{equation}
n^N_{ex} = {2 \over {3\pi^2}} \left((\Delta + k_F)^3 - k_F^3 \right),
\end{equation}
where we have assumed the thickness of the shell is $\Delta$, and there is one flavor and two spins of nucleons.
 The baryon number Fermi momentum  $k_F$.  The top of the quark Fermi sea is $k_Q = k_F/N_c$, where $N_c$ is the number of colors.  The bottom of the nucleon shell is $k_F$.
We will for the time being assume non-relativistic quarks and nucleons so that the kinetic energy contribution to the energy density from the shell is
\begin{equation}
\epsilon^N_{ex} = {1 \over {5 \pi^2}} {1 \over {M}} \left( (\Delta+k_F)^5 - k_F^5 \right).
\end{equation}

The quark contribution to the baryon number density for a free gas of quarks is 
\begin{equation}
 n_{Q }^N= {2 \over {3\pi^2}}{1 \over N_c^3} k_F^3.
\end{equation}
Unfortunately, at small $k_Q$ as the quark density  becomes non-zero, there is too rapid a rise with $k_Q$ to generate a physically acceptable sound velocity.  We had to modify the very low density quark Fermi distribution in a way so that we maintain the proper behaviour for large Fermi momenta, $k_Q >> \Lambda_{\textrm{QCD}}$,
and that then the density is that for free particles.  For small $k_Q$, we make the density approach zero as $k_Q^2$.  This effectively increases the density of states for very low quark densities.  In this region of course non-perturbative effects are important, and it is difficult to a priori determine the relationship between density and quark Fermi momentum, so this modification should be taken as a reasonable guess that
allows for a non-singular transition to Quarkyonic Matter.  We take
\begin{equation}
  \tilde{n}_Q^N = {2 \over {3\pi^2}} \left( (k_Q^2 + \Lambda^2)^{\frac{3}{2}}- \Lambda^3 \right). \label{qir}
\end{equation}
This corresponds to modifying the density of states by $1 \rightarrow \sqrt{k_Q^2 + \Lambda^2}/k_Q$.  Here $\Lambda$ is a 
parameter of order $\Lambda \sim \Lambda_{\textrm{QCD}}$

The equation for the baryon number density lets us compute $\Delta$ as
\begin{equation}
  \Delta = \left(\frac{3\pi^2}{2} n_{ex}^N +k_F^3\right)^{1/3} - k_F.
\end{equation}
We will analytically derive relations for the energy density in the nonrelativstic limit.  (In the numerical results we soon present,
we will use relativistic expression for the energy densities.) 
The nucleon energy density is
\begin{equation}
\epsilon^{N}_{ex} = {1 \over {5\pi^2}} {1 \over {M}} \left( \left(\frac{3\pi^2}{2} n_{ex}^N + k_F^3 \right)^{\frac{5}{3}} - k_F^{5} \right),
\end{equation}
with 
\begin{equation}
  \epsilon_N^N = \epsilon_{ex}^N V_{ex}/V = \epsilon_{ex}^N (1-n_N^N/n_0).
\end{equation}
For the quarks, we find, using the modified density of states,
\begin{equation}
  \epsilon_Q^N = {N_c^2 \over \pi^2} {1 \over {M}} \left\{ {1 \over 5} (k_Q^2 + \Lambda^2)^{5/2}  - {1 \over 3} \Lambda^2 ( k_Q^2 + \Lambda^2)^{3/2}  +{2 \over 15} \Lambda^5  \right\}.
\end{equation}
Note that  when the quark density is of the order of the nucleon density, then kinetic energies are of comparable magnitude.

We must extremize the energy density 
\begin{equation}
  \epsilon = \epsilon_Q^N + \epsilon_N^N,
  \end{equation}
  with respect to quark and nucleon densities
subject to the constraint that the total baryon number is held fixed.

We now compute the relativistc expressions for the energy density.  We first do this for the case of the naive free particle density of states,
and then later modify the density of qaurks states as described above. 
For the relativistic case with ideal gas of quarks, the energy density can be summarized as
\begin{align}
\epsilon &= 4 \left( 1- \frac{n_N^N}{n_0} \right)\int_{k_F}^{k_F+\Delta} \frac{d^3 k}{(2\pi)^3} \left((N_c m_Q)^2+k^2 \right)^{\frac{1}{2}}+4N_c \int^{k_F/N_c}_0 \frac{d^3 k}{(2\pi)^3} \left(m_Q^2+k^2 \right)^{\frac{1}{2}},\label{eden}
\end{align}
where flavor symmetry is considered. Minimizing the energy density functional with respect to nucleon density for a fixed total baryon density at $dn_N=dn_Q+dn_N^N=0$ we can obtain the energy density at the minimum.
%\begin{align}
%N_c& \left(\left(\frac{3\pi^2}{2}  n_Q \right)^{\frac{2}{3}}+m_Q^2\right)^{\frac{1}{2}} \nonumber\\
%&= -\frac{1}{n_0} \frac{2}{\pi^2} \int_{k_F}^{k_F+\Delta} dk k^2 ((N_c m_Q)^2+k^2)^{\frac{1}{2}} \nonumber\\
%&~~+ \left( 1- \frac{n_N^N}{n_0} \right) \left( \left(\left(\frac{3\pi^2}{2}(n_{ex} + N_c^3 n_Q)\right)^{\frac{2}{3}}+(N_c m_Q)^2\right)^{\frac{1}{2}}\left(\left( 1- \frac{n_N^N}{n_0} \right)^2  -N_c^3\right)+ \left(\left(\frac{3\pi^2}{2}N_c^3 n_Q\right)^{\frac{2}{3}}+(N_c m_Q)^2\right)^{\frac{1}{2}} N_c^3 \right).\label{const1}
%\end{align}
The minimum of the energy density functional at a certain baryon density is the energy density at that baryon density. The chemical potential $\mu$ then can be found  as:
\begin{align}
\mu=\frac{\partial \epsilon_{\text{min}}}{\partial n_N},
\end{align}
%\begin{align}
%\mu=\frac{\partial \epsilon_{\text{min}}}{\partial n_N} = -\frac{1}{n_0} \frac{2}{\pi^2} \int_{k_F}^{k_F+\Delta} dk k^2 (( N_c m_Q)^2+k^2)^{\frac{1}{2}}+\left( 1- \frac{n_N^N}{n_0} \right)^{-1}    \left(\left(\frac{3\pi^2}{2}(n_{ex} + N_c^3 n_Q)\right)^{\frac{2}{3}}+(N_c m_Q)^2\right)^{\frac{1}{2}},
%\end{align}
where $\epsilon_{\text{min}}$ is the energy density of the baryons at baryon density $n_N$ obtained though the minimization procedure outlined above. 

Unfortunately, for an ideal gas of quarks, the behaviour of the energy density as $k_F^5 \sim n_Q^{5/3}$ generates a singularity when one computes the sound velocity, that involve two derivatives with respect to $n_Q$.  As described above, we need to slow the rapid turn on of the quark density as a function of $k_Q$ as described above. 

\begin{figure}
\includegraphics[width=0.45\textwidth]{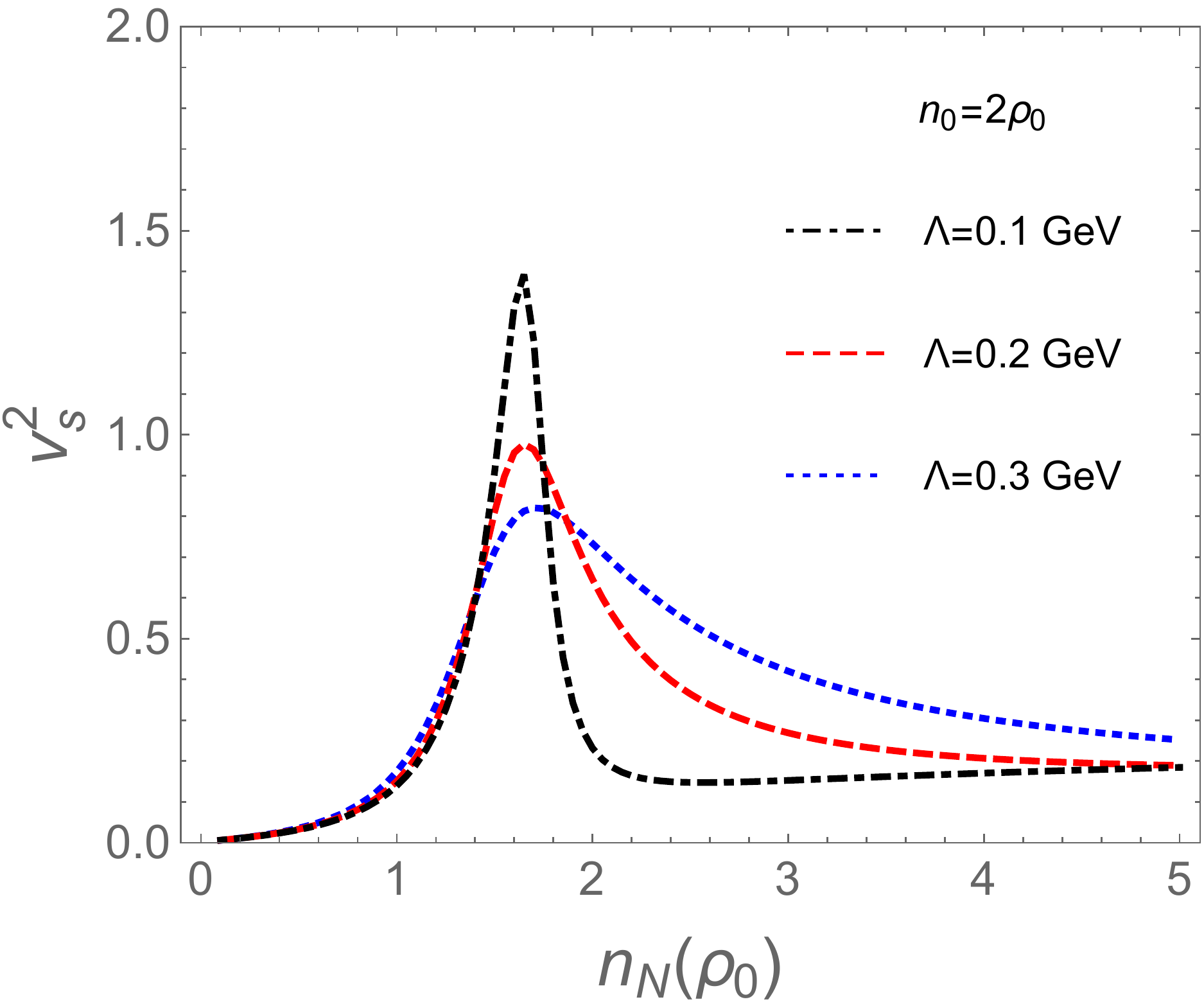}
\includegraphics[width=0.45\textwidth]{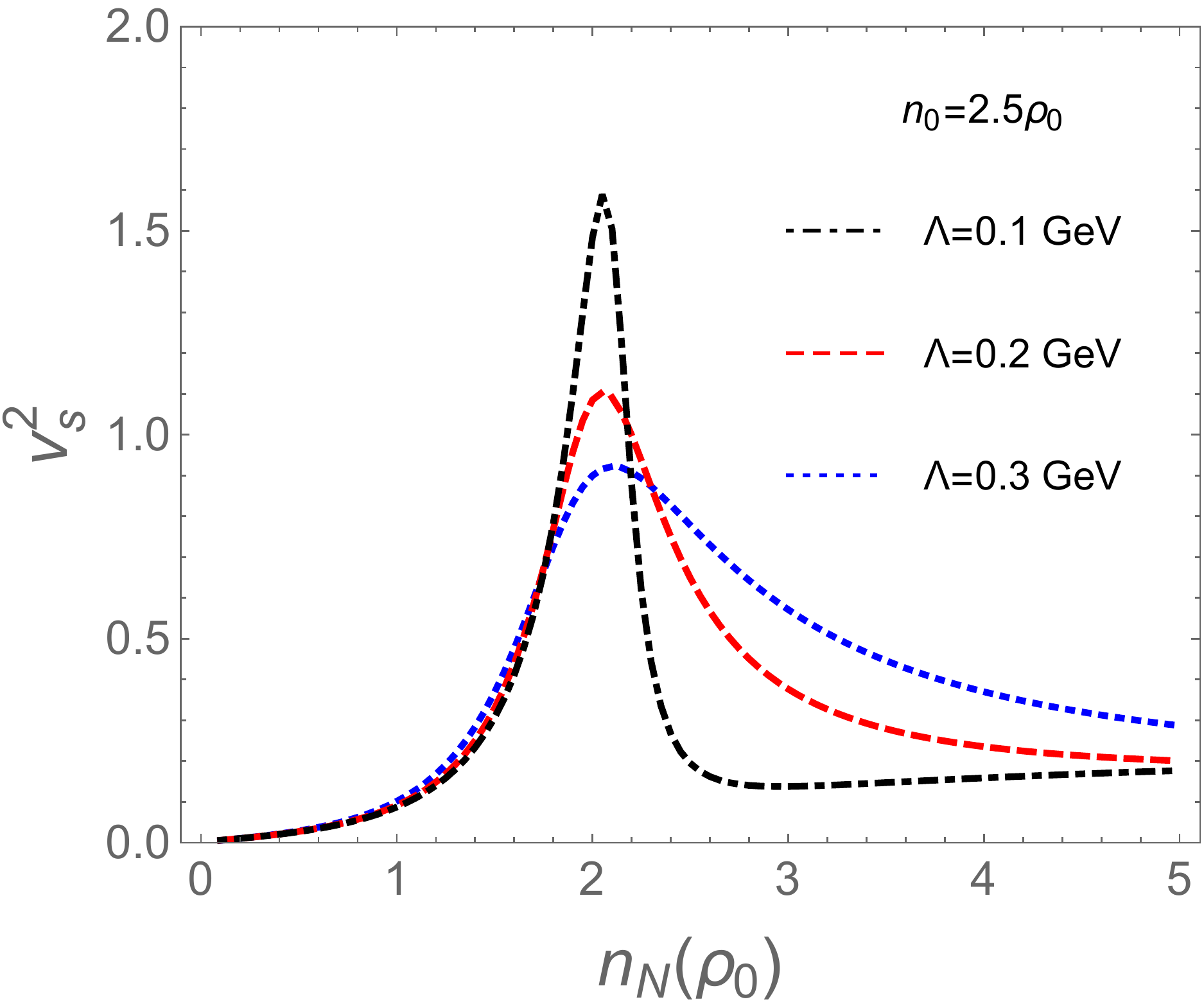}
\caption{$v_s^2$ with different non-perturbative IR scale $\Lambda$s for quark degree of freedom ($N_c=3$). Each color corresponds to $\Lambda = 0.3~\textrm{GeV}$ (blue), $\Lambda = 0.2~\textrm{GeV}$ (red), and $\Lambda = 0.1~\textrm{GeV}$ (black). The hard core density is set as $n_0=2.0\rho_0$ for left and $n_0=2.5\rho_0$ for right.} \label{f5}
\end{figure}

\begin{figure}
\includegraphics[width=0.45\textwidth]{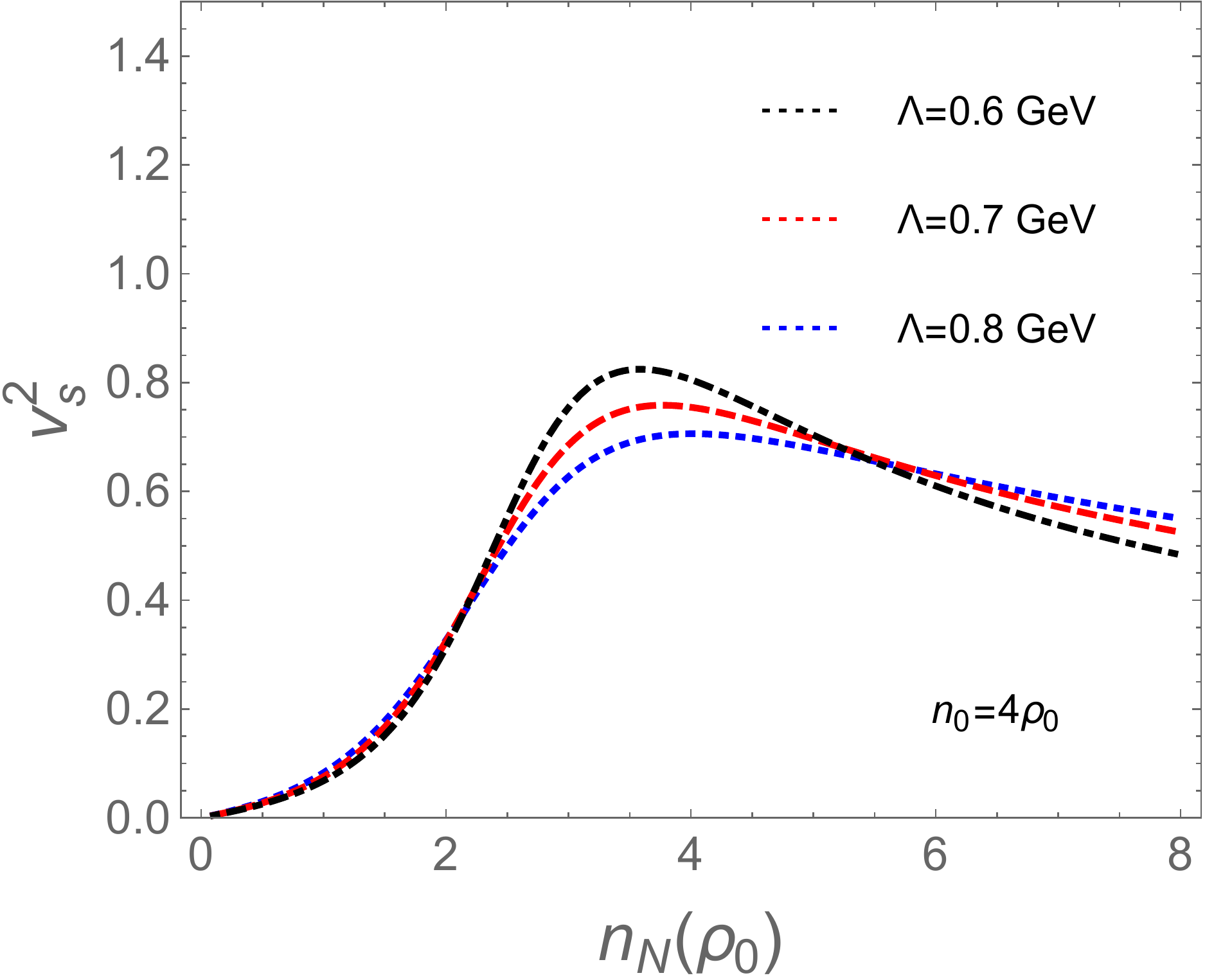}
\includegraphics[width=0.45\textwidth]{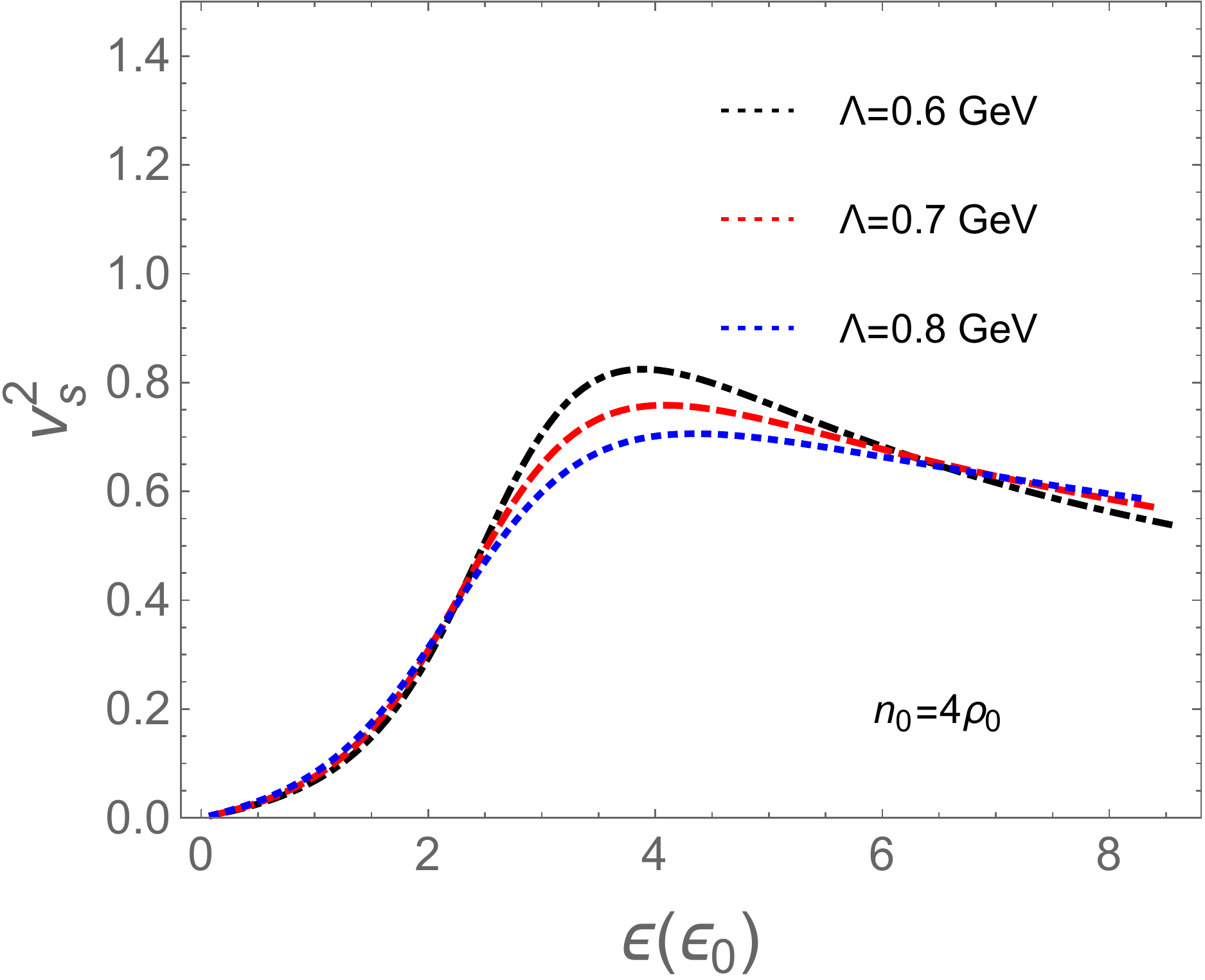} 
\caption{$v_s^2$ with different non-perturbative IR scale $\Lambda$s for quark degree of freedom ($N_c=3$). Each left and right graph is plotted in terms of the total baryon number density and total energy density, respectively. Each color corresponds to $\Lambda = 0.6~\textrm{GeV}$ (black), $\Lambda = 0.7~\textrm{GeV}$ (red), and $\Lambda = 0.8~\textrm{GeV}$ (blue). The hard core density is set as $n_0=4.0\rho_0$.} \label{f6}
\end{figure}

If one assigns some non-perturbative gas behavior as considered in Eq.~\eqref{qir}, the energy density~\eqref{eden} can be redefined as
\begin{align}
\tilde{\epsilon} &= 4 \left( 1- \frac{n_N^N}{n_0} \right)\int_{k_F}^{k_F+\Delta} \frac{d^3 k}{(2\pi)^3} \left((N_c m_Q)^2+k^2 \right)^{\frac{1}{2}}+ \frac{2 Nc}{\pi^2} \int^{k_F/N_c}_0  d k k \left(\Lambda^2+k^2 \right)^{\frac{1}{2}} \left(m_Q^2+k^2 \right)^{\frac{1}{2}}. 
\end{align}
Again, minimizing the energy functional $\tilde{\epsilon}$ we can obtain the energy density as a function of the baryon density as well as the corresponding chemical potential and speed of sound.
The sound veoicity is illustrative and contains qualitative information about how
the stiffness of the equation of state depends upon hard core density and our scale $\Lambda$.
The squared speed of sound is plotted in Fig.~\ref{f5} and Fig. ~\ref{f6}. If one considers non-perturbative IR scale $\Lambda=0.3~\textrm{GeV}$ with  $n_0=2.5\rho_0$, the sound velocity  rapidly increases to a maximum which is also approximately the value at the  hard core density  near the onset of the quark degrees of freedom and becomes moderate after the onset density.   Fig.~\ref{f6} is at a somewhat more realistic core density with  more realistic choices for $\Lambda$ which might be more applicable for neutron star phenomenology. We have to choose 
$\Lambda$ so that the sound velocity is in a reasonable range, a higher value of the core density shifts the maximum to a higher value of $n_N \sim n_0$. The maximum value of the sound velocity increase as the scale $\Lambda$ decreases.  $\Lambda$ must be chosen sufficiently large to require $v_s^2 < 1$.

%\begin{figure}
%\includegraphics[width=0.45\textwidth]{fig3a.pdf}
%\includegraphics[width=0.45\textwidth]{fig3b.pdf}
%\caption{$ \partial \tilde{n}_Q / \partial n_N$ with $\Lambda=0.1~\textrm{GeV}$ for left and $\Lambda=0.3~\textrm{GeV}$ for right graph ($N_c=3$). The %hard core density is set as $n_0=2.5 \rho_0$.} \label{f3}
%\end{figure}

%\begin{figure}
%\i%ncludegraphics[width=0.45\textwidth]{fig4a.pdf}
%\includegraphics[width=0.45\textwidth]{fig4b.pdf}  
%\caption{Chemical potential (left) and $v_s^2$ (right) with non-perturbative IR scale $\Lambda$ for quark degree of freedom ($N_c=3$). The hard core %density is set as $n_0=2.5\rho_0$ and $\Lambda=0.3~\textrm{GeV}$.} \label{f4}
%\end{figure}

\section{The Shell Thickness in the Thin Shell Limit}

The shell becomes thin in the region where we can treat the quarks and nucleon as non relativistic, but where $k_f^3 >> 3\pi^2 n^N_{ex}$.  Let us evaluate the shell thickness using the naive expression for the quark density $\Lambda =0$.  
The absolute minimum for $n_N^N$ for $n_N^N$ is close to $n_0$.
This is the limit where $\Delta << k_F$ corresponding to a thin shell for the baryons.  As one makes the transition between Nucleon Matter and Quarkyonic Matter, that is, as the shell begins to form from a Fermi sphere of nucleon and when the Fermi sphere of quarks forms, the situation is more complicated, and we will rely on numerical analysis to handle this region.    To this end we compute the energy density when $n_{ex} << k_F^3$.   

First let us compute the energy density of the quarks:
\begin{equation}
  \epsilon_Q^N = {3 \over 5} n_Q^N {k_F^2 \over {2M}} = {3 \over {10M}} \left(\frac{3\pi^2}{2} \right)^{2/3} N_c^2 (n_Q^N)^{5/3} = \kappa N_c^2 (n_Q^N)^{5/3}.
\end{equation}
Using  $\kappa = {3 \over {10M}}\left(3\pi^2/2 \right)^{2/3} $, we writre $n_Q^N = n_N - n_N^N$ and expand the previous equation to second order in $n_N^N$,
\begin{equation}
  \epsilon_Q^N = \kappa  N_c^2  n_N^{5/3} \left(1 -{{5n_N^N} \over {3n_N}} + {5 \over 9} \left( {n_N^N \over n_N} \right)^2 + \cdots \right).\label{energydq}
\end{equation}

Now let us compute the nucleon energy to a simlar accuracy
\begin{equation}
  \epsilon_N^N ={5 \over 3}  \kappa  N_c^2 n_N^N (n^N_Q)^{2/3} \left\{ 1 + {n^{N}_{ex} \over {3 N_c^3 n_Q^N}}  \right\}.\label{energydn}
\end{equation}

For small $n_N^N << n_N$ as is true near zero nucleon density, the total energy density is approximately:
\begin{equation}
  \epsilon_N \sim \kappa N_c^2 n_N^{5/3} \left\{1 - {5 \over 9} \left( {n_N^N \over n_N} \right)^2 \left(\frac{N_c^3-1}{N_c^3}\right)\right\}.
\end{equation}
This has a maximum near $n_N^N \rightarrow 0$.

Now let us analyze near the minimum which occurs for $n_N^N$ near to the hard core density $n_0$.  I will consider the case  where $n_N >> n_0$.
In this case there is a singular contribution from the second term in Eq.~\eqref{energydn}, that can compensate for the first term in this equation  and the contrbution from the quark energy density. We will evaluate this contribution and show that higher order terms in the expansion of Eq.~\eqref{energydq} may be ingnored, that is we are in the thin shell limit.
First we evaluate
\begin{equation}
  \epsilon_Q^N \sim \kappa N_c^2 (n_N - n_N^N)^{5/3} \sim \kappa N_c^2 n_N^{5/3} \left(1 -{{5n_N^N} \over {3n_N}} + {5 \over 9} \left( {n_N^N \over n_N} \right)^2 \right).
\end{equation}
Similarly,
\begin{equation}
   \epsilon_N^N \sim   {5 \over 3} \kappa N_c^2 n_N^N n_N^{2/3} \left( 1 -{2\over 3} {n_N^N \over n_N}+{ n_N^N \over {3N_c^3n_N(1-n_N^N/n_0)}} \right).
\end{equation}
When the energy density is added the leading order terms cancel.  The minimum is found when we extremize 
\begin{equation} 
 \epsilon_N -\kappa N_c^2 n_N^{5/3} \sim {5 \over 9} \kappa N_c^2 n_N^{5/3} \left( {n_N^N  \over n_N} \right)^2 \left\{   -1 + {1 \over {N_c^3(1-n_N^N/n_0)} } \right\}.
\end{equation}
This has an extremum when $n/n_0 = 1- b N_c^{-3/2}$ and is independent of density.  The minimum occurs
when
\begin{equation}
  n_N^N/n_0 = 1- {1 \over \sqrt{2N_c^3}} \sim 0.86,
\end{equation}
in agreement with our numerical analysis.

The thickness of the shell is approximately,
\begin{equation}
  \Delta \sim n^{N}_{ex} /k_F^2 \sim {n_0 \over {N_c^{1/2} k_Q^2}}.
\end{equation}
This equation implies a rapid transition to a thin shell in the region where the matter is transitioning from primarily nucleons to primarily
quarks. The precise value to the thickness will of course depend upon the details of the relationship between quark Fermi momentum and density, and this is of course corrected by relativistic and non-perturbative effects in the transition region.  It is nevertheless encouraging to see the overall $1/k_F^2$ dependence of the thickness as predicted in Ref.~\cite{McLerran:2018hbz}

\begin{figure}
\begin{center}
\includegraphics[width=0.45\textwidth]{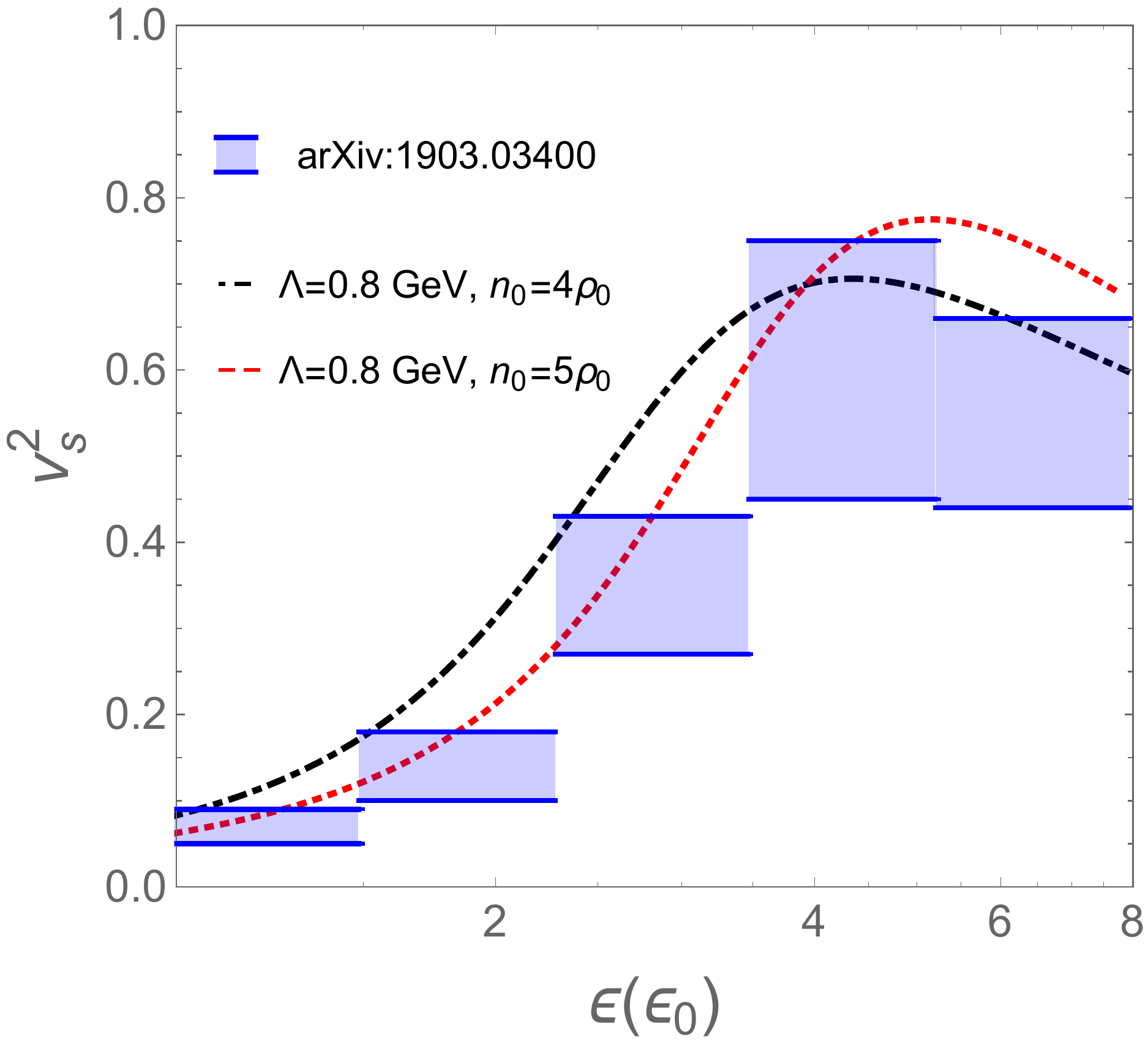}
\end{center}
\caption{The sound velocity for a 2 flavor equation of state as above with value of  $v_s^2$ vs energy desnity $\epsilon$ compared to the result of Ref.~\cite{Fujimoto:2019hxv} (blue boxes).} \label{kenji}
\end{figure}

\section{Summary}

A comparison with possible acceptable neutron star equations of state is beyond what we present here.  Such a comparison would require that we properly match onto well determined nuclear matter equations of states at 1-2 times nuclear density, and that we include effects of beta equilibrium in the equation of state.  Nevertheless, we can take our naive equations of state computed above, and see if for chosen values of the hard core density, we are in approximately the right range to describe equations of state extracted from the properties of neutron stars.  Such a comparison to the extraction of Fujimoto et.al.~\cite{Fujimoto:2019hxv} is shown in Fig.~\ref{kenji}.  This plot shows semi-quantitative agreement with the extracted equation of state. Interestingly, our results have commonalities with the results of \cite{Paeng:2017qvp} where the authors consider skyrmion description of compressed baryonic matter.

\section{Acknowledgements}  All three authors would like to acknowledge very useful conversations with Sanjay Reddy, and Krzysztof Redlich.  L. McLerran acknowledges the kind hospitality of Young Lian Ma and Manque Rho at the Workshop on New Staes of Dense Matter at Jilin University where part of this work was completed. The work of Kie Sang Jeong was partially supported by the NRF of Korea under Grant No.
NRF-2017R1D1A1B03033685 and Basic Science
Research Program through the NRF funded by the Ministry of Science, ICT and Future Planning.
 Kie Sang Jeong and Srimoyeee Sen acknowledge the support of the Simons Foundation under the Multifarious Minds Program grant . 
The work of Larry McLerran and Srimoyee Sen was supported by the U.S. DOE under Grant No. DE-FG02- 00ER41132.


\begin{thebibliography}{70}

%\cite{McLerran:2007qj}
\bibitem{McLerran:2007qj} 
  L.~McLerran and R.~D.~Pisarski,
  %``Phases of cold, dense quarks at large N(c),''
  Nucl.\ Phys.\ A {\bf 796}, 83 (2007)
  doi:10.1016/j.nuclphysa.2007.08.013
  [arXiv:0706.2191 [hep-ph]].
  %%CITATION = doi:10.1016/j.nuclphysa.2007.08.013;%%
  %517 citations counted in INSPIRE as of 13 Dec 2018
	
	%\cite{Fukushima:2015bda}
\bibitem{Fukushima:2015bda} 
  K.~Fukushima and T.~Kojo,
  %``The Quarkyonic Star,''
  Astrophys.\ J.\  {\bf 817}, no. 2, 180 (2016)
  doi:10.3847/0004-637X/817/2/180
  [arXiv:1509.00356 [nucl-th]].
  %%CITATION = doi:10.3847/0004-637X/817/2/180;%%
  %39 citations counted in INSPIRE as of 01 Oct 2019	
	
	%\cite{DiToro:2009ig}
\bibitem{DiToro:2009ig} 
  M.~Di Toro, B.~Liu, V.~Greco, V.~Baran, M.~Colonna and S.~Plumari,
  %``Symmetry Energy Effects on the Mixed Hadron-Quark Phase at High Baryon Density,''
  Phys.\ Rev.\ C {\bf 83}, 014911 (2011)
  doi:10.1103/PhysRevC.83.014911
  [arXiv:0909.3247 [nucl-th]].
  %%CITATION = doi:10.1103/PhysRevC.83.014911;%%
  %35 citations counted in INSPIRE as of 12 Aug 2019
	
%\cite{Hebeler:2013nza}
\bibitem{Hebeler:2013nza} 
  K.~Hebeler, J.~M.~Lattimer, C.~J.~Pethick and A.~Schwenk,
  %``Equation of state and neutron star properties constrained by nuclear physics and observation,''
  Astrophys.\ J.\  {\bf 773}, 11 (2013)
  doi:10.1088/0004-637X/773/1/11
  [arXiv:1303.4662 [astro-ph.SR]].
  %%CITATION = doi:10.1088/0004-637X/773/1/11;%%
  %289 citations counted in INSPIRE as of 12 Aug 2019
 %\cite{McLerran:2018hbz}
\bibitem{McLerran:2018hbz} 
  L.~McLerran and S.~Reddy,
  %``Quarkyonic Matter and Neutron Stars,''
  arXiv:1811.12503 [nucl-th].
  %%CITATION = ARXIV:1811.12503;%%
  
  %\cite{Masuda:2012ed}
	%\cite{Philipsen:2019qqm}
\bibitem{Philipsen:2019qqm} 
  O.~Philipsen and J.~Scheunert,
  %``QCD in the heavy dense regime for general $N_c$: On the existence of quarkyonic matter,''
  arXiv:1908.03136 [hep-lat].
  %%CITATION = ARXIV:1908.03136;%%
	%\cite{Han:2019bub}
\bibitem{Han:2019bub} 
  S.~Han, M.~A.~A.~Mamun, S.~Lalit, C.~Constantinou and M.~Prakash,
  %``Treating quarks within neutron stars,''
  arXiv:1906.04095 [astro-ph.HE].
  %%CITATION = ARXIV:1906.04095;%%
  %1 citations counted in INSPIRE as of 12 Aug 2019
	
	%\cite{Demorest:2010bx}
\bibitem{Demorest:2010bx} 
  P.~Demorest, T.~Pennucci, S.~Ransom, M.~Roberts and J.~Hessels,
  %``Shapiro Delay Measurement of A Two Solar Mass Neutron Star,''
  Nature {\bf 467}, 1081 (2010)
  doi:10.1038/nature09466
  [arXiv:1010.5788 [astro-ph.HE]].
  %%CITATION = doi:10.1038/nature09466;%%
  %2018 citations counted in INSPIRE as of 12 Aug 2019
	
	%\cite{Antoniadis:2013pzd}
\bibitem{Antoniadis:2013pzd} 
  J.~Antoniadis {\it et al.},
  %``A Massive Pulsar in a Compact Relativistic Binary,''
  Science {\bf 340}, 6131 (2013)
  doi:10.1126/science.1233232
  [arXiv:1304.6875 [astro-ph.HE]].
  %%CITATION = doi:10.1126/science.1233232;%%
  %1111 citations counted in INSPIRE as of 12 Aug 2019
	
\bibitem{Masuda:2012ed} 
  K.~Masuda, T.~Hatsuda and T.~Takatsuka,
  %``Hadron?quark crossover and massive hybrid stars,''
  PTEP {\bf 2013}, no. 7, 073D01 (2013)
  doi:10.1093/ptep/ptt045
  [arXiv:1212.6803 [nucl-th]].
  %%CITATION = doi:10.1093/ptep/ptt045;%%
  %79 citations counted in INSPIRE as of 28 Jun 2019
   
  %\cite{Kojo:2014rca}
\bibitem{Kojo:2014rca} 
  T.~Kojo, P.~D.~Powell, Y.~Song and G.~Baym,
  %``Phenomenological QCD equation of state for massive neutron stars,''
  Phys.\ Rev.\ D {\bf 91}, no. 4, 045003 (2015)
  doi:10.1103/PhysRevD.91.045003
  [arXiv:1412.1108 [hep-ph]].
  %%CITATION = doi:10.1103/PhysRevD.91.045003;%%
  %70 citations counted in INSPIRE as of 28 Jun 2019
  
  %\cite{Bedaque:2014sqa}
\bibitem{Bedaque:2014sqa} 
  P.~Bedaque and A.~W.~Steiner,
  %``Sound velocity bound and neutron stars,''
  Phys.\ Rev.\ Lett.\  {\bf 114}, no. 3, 031103 (2015)
  doi:10.1103/PhysRevLett.114.031103
  [arXiv:1408.5116 [nucl-th]].
  %%CITATION = doi:10.1103/PhysRevLett.114.031103;%%
  %66 citations counted in INSPIRE as of 28 Jun 2019

%\cite{Ma:2018qkg}
\bibitem{Ma:2018qkg} 
  Y.~L.~Ma and M.~Rho,
  %``Sound velocity and tidal deformability in compact stars,''
  arXiv:1811.07071 [nucl-th].
  %%CITATION = ARXIV:1811.07071;%%
  %5 citations counted in INSPIRE as of 28 Jun 2019


  %\cite{Tews:2018kmu}
\bibitem{Tews:2018kmu} 
  I.~Tews, J.~Carlson, S.~Gandolfi and S.~Reddy,
  %``Constraining the speed of sound inside neutron stars with chiral effective field theory interactions and observations,''
  Astrophys.\ J.\  {\bf 860}, no. 2, 149 (2018)
  doi:10.3847/1538-4357/aac267
  [arXiv:1801.01923 [nucl-th]].
  %%CITATION = doi:10.3847/1538-4357/aac267;%%
  %46 citations counted in INSPIRE as of 28 Jun 2019

  
  %\cite{Vuorinen:2018qzx}
\bibitem{Vuorinen:2018qzx} 
  A.~Vuorinen,
  %``Neutron stars and stellar mergers as a laboratory for dense QCD matter,''
  Nucl.\ Phys.\ A {\bf 982}, 36 (2019)
  doi:10.1016/j.nuclphysa.2018.10.011
  [arXiv:1807.04480 [nucl-th]].
  %%CITATION = doi:10.1016/j.nuclphysa.2018.10.011;%%
  %6 citations counted in INSPIRE as of 28 Jun 2019
  
  %\cite{Fujimoto:2019hxv}
\bibitem{Fujimoto:2019hxv} 
  Y.~Fujimoto, K.~Fukushima and K.~Murase,
  %``Mapping neutron star data to the equation of state of the densest matter using the deep neural network,''
  arXiv:1903.03400 [nucl-th].
  %%CITATION = ARXIV:1903.03400;%%
  %2 citations counted in INSPIRE as of 28 Jun 2019
  
  %\cite{Zalewski:2015yea}
\bibitem{Zalewski:2015yea} 
  K.~Redlich and K.~Zalewski,
  %``Thermodynamics of the low density excluded volume hadron gas,''
  Phys.\ Rev.\ C {\bf 93}, no. 1, 014910 (2016)
  doi:10.1103/PhysRevC.93.014910
  [arXiv:1507.05433 [hep-ph]].
  %%CITATION = doi:10.1103/PhysRevC.93.014910;%%
  %7 citations counted in INSPIRE as of 01 Jul 2019
  
  %\cite{Redlich:2016dpb}
\bibitem{Redlich:2016dpb} 
  K.~Redlich and K.~Zalewski,
  %``Thermodynamics of Van der Waals Fluids with quantum statistics,''
  Acta Phys.\ Polon.\ B {\bf 47}, 1943 (2016)
  doi:10.5506/APhysPolB.47.1943
  [arXiv:1605.09686 [cond-mat.quant-gas]].
  %%CITATION = doi:10.5506/APhysPolB.47.1943;%%
  %17 citations counted in INSPIRE as of 01 Jul 2019
  
  %\cite{Vovchenko:2015vxa}
\bibitem{Vovchenko:2015vxa} 
  V.~Vovchenko, D.~V.~Anchishkin and M.~I.~Gorenstein,
  %``Van der Waals Equation of State with Fermi Statistics for Nuclear Matter,''
  Phys.\ Rev.\ C {\bf 91}, no. 6, 064314 (2015)
  doi:10.1103/PhysRevC.91.064314
  [arXiv:1504.01363 [nucl-th]].
  %%CITATION = doi:10.1103/PhysRevC.91.064314;%%
  %33 citations counted in INSPIRE as of 01 Jul 2019

%\cite{Paeng:2017qvp}
\bibitem{Paeng:2017qvp} 
  W.~G.~Paeng, T.~T.~S.~Kuo, H.~K.~Lee, Y.~L.~Ma and M.~Rho,
  %``Scale-invariant hidden local symmetry, topology change, and dense baryonic matter. II.,''
  Phys.\ Rev.\ D {\bf 96}, no. 1, 014031 (2017)
  doi:10.1103/PhysRevD.96.014031
  [arXiv:1704.02775 [nucl-th]].
  %%CITATION = doi:10.1103/PhysRevD.96.014031;%%
  %23 citations counted in INSPIRE as of 01 Oct 2019
\end{thebibliography}
\end{document}